\documentstyle[epsfig,amsmath]{europhys}

\def\etal{{\hbox{{\it\ et al.\/}\rm :\ }}}

\def\And{{\rm and\ }}

\def\stars{\bigskip\centerline{***}\medskip}

\newif\ifboo \boofalse

\begin{document}

\euro{}{}{}{}

\Date{}

\shorttitle{J. BADRO \etal MELTING AND PRESSURE-INDUCED
AMORPHIZATION OF QUARTZ}

\title{Melting and Pressure-Induced Amorphization of Quartz}

\author{James Badro\inst{1}, Philippe Gillet\inst{1} \And
Jean-Louis Barrat\inst{2}}

\institute{
     \inst{1} Laboratoire de Sciences de la Terre ---
\'{E}cole normale sup\'{e}rieure de Lyon\\
46, all\'{e}e d'Italie 69364 Lyon cedex 07, France.\\
     \inst{2} D\'epartement de Physique des Mat\'eriaux ---
Universit\'e Claude Bernard -- Lyon I\\
43, bd. du 11 novembre 1918, 69622 Villeurbanne cedex, France.
}

\rec{}{}

\pacs{
\Pacs{62}{50.+p}{High Pressure}
\Pacs{61}{20.-p}{Structure of Liquids}
\Pacs{64}{70.Dv}{Solid--Liquid Transitions}
      }

\maketitle

\begin{abstract}
It has recently been shown that amorphization and melting of ice
\cite{mishima2} were intimately linked. In this letter, we infer from
molecular dynamics simulations on the SiO$_{2}$ system that the
extension of the $\alpha -$quartz melting line in the metastable
pressure--temperature domain is the pressure-induced amorphization
line. It seems therefore likely that melting is the physical phenomenon
responsible for pressure induced amorphization. Moreover, we show that
the  structure of a "pressure glass" is similar to that of a very
rapidly ($10^{13}$ to $10^{14}$ kelvins per second) quenched thermal
glass.
\end{abstract}

\section{Introduction}

Pressure induced amorphization (PIA) is an  intriguing phenomenon. The
first report of this transition goes back to 1984 when Mishima {\em et al.\/}
\cite{mishima1} found that ice--Ih cooled down to 77~K and pressurized to
1~GPa failed to transform into crystalline Ice--IX and became amorphous.
Since then, a large number of materials were found to undergo similar
transitions \cite{richet2} at different pressures and temperatures. In
particular, quartz, one of the  most studied minerals, was found to
amorphize \cite{hemley1} at room temperature at pressures between 15 and 25~GPa.
A series of experiments \cite{meade1, itie1} and molecular dynamics
(MD) simulations \cite{tse1,tse2,chelikowsky1} have shown that this
transition is accompanied by an increase in silicon coordination.
However, the thermodynamic nature of this transition remains very
poorly defined. It has been argued on the basis of lattice dynamics
calculations using the quasi-harmonic approximation and the Born
stability criteria \cite{binggeli3} that the PIA of $\alpha -$quartz
took place as the result of an elastic instability \cite{mcneil1}
occurring at approximately 22~GPa. More recently, it has been proposed
that a dynamical instability occured at lower pressure \cite{watson1}
(18.5~GPa) than the elastic instability, and that the sizes and
geometries of the simulation boxes dramatically affected the existence
and thresholds of the reported instabilities. Results obtained by {\em
ab initio} calculations within the local density approximation of the
density functional theory showed \cite{binggeli97} that the elastic
instability occured at 29~GPa, but all recent MD
calculations \cite{tse1,tse2,badro96} suggest that amorphization
occurs at 22~GPa, independently
of the size or geometry of the simulation box, so that the role of the
instability in the pressure induced amorphization process is still a
matter of debate \cite{watson1}.

Hemley {\em et al.\/} \cite{hemley2} proposed that PIA of quartz
represents the metastable extension of the melting curve of SiO$_2$.
Richet \cite{richet1} using thermodynamic arguments has inferred that
the vitrification upon room temperature {\em decompression} or upon
room pressure heating of some high pressure silicate minerals
represents crystal fusion. In the first report on the PIA of ice,
Mishima proposed that the transformation could be related to a
metastable melting process, and the author recently showed \cite{mishima2}
that the pressure-induced transformation towards amorphous ice
(high density amorphous phase) can be described as a melting transition
to an extremely unrelaxed amorphous state. This state then slowly
relaxes towards an energetically more favourable glass-like state.
It has been shown using Raman spectroscopy that the amorphization
induced by heating of stishovite leads to a highly unrelaxed glassy
state \cite{gillet90,richet2,grimsditch94} exhibiting the features of
the pressure glass obtained upon room temperature compression and
decompression of quartz\footnote{the reference pressure glass we have
modelized in this study}, which then gradually transforms into a
relaxed glassy state indistinguishable (at least from the vibrational
point of view) from the usual thermal glass, thus indicating that the
two "glassy" states probably belong to the same basin in configurational
space.

In this letter, we present results from MD
simulations that provide new insights into the relationship between PIA
and melting, and report the structural and thermodynamic similarities
between amorphous SiO$_2$ prepared by static compression of quartz at
room temperature and quenched silica glass.

\section{Results and Discussion}

MD simulations have been used to investigate the melting and
amorphization curves of $\alpha -$quartz in the metastable P--T domain
in which the nucleation of the high pressure polymorphs cannot be bypassed
in laboratory experiments. Also, the cooling rates used for quenching
liquid SiO$_2$ in MD simulations are by far larger than those used by
current experimental techniques.
MD calculations were carried out in an (N,P,T) isothermal-isobaric
ensemble, for a system of  480 SiO$_2$ molecules.  The BKS
\cite{beest1} pairwise interaction potential which has proven to
provide a very accurate description of silica at high pressure
\cite{tse1, badro96, tse2}, was used.  A system in the $\alpha -$quartz
structure was subjected to compression at different temperatures, or to
isobaric heating, until either amorphization or melting was observed.
In the low-temperature high-pressure part of the phase diagram relative
to non-equilibrium melting, these amorphous states were detected by a
combined analysis of the radial distribution function and the energy
jump at the transition, along with an analysis of the average positions
of the particles integrated on the duration of the simulation. In the
high-temperature low-pressure domain relative to the equilibrium
melting curve, and thanks to the power of modern parallel computers, we
could achieve a real thermodynamic determination of the melting points
within reasonable computation time, which was impossible to carry out
only a few years ago on such viscous systems \cite{barrat2}.  Instead of
calculating the differences of free enthalpy between the solid and
liquid phases, we used an equivalent technique that consisted in
performing two phase molecular dynamics simulations, and monitored the
energy variation of the system with time, along with the relative
increase or decrease in one of the two phases in the simulation box.
With computation times of 0.3 ns, it was possible to determine the
theoretical melting curve for the model, with an accuracy of the
order of 50~K at 10 and 5~GPa.

\begin{figure}[h]
\centerline{\epsfig{file=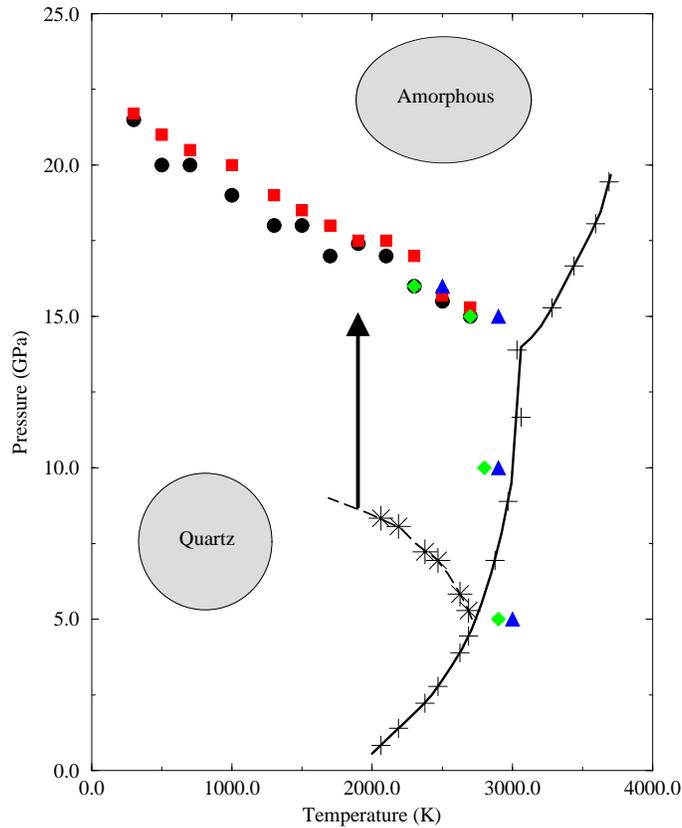,width=9cm}}
\vspace{1cm}
\caption{The quartz-melt coexistence curve (full symbols). This
line separates the phase diagram into two distinct areas: the quartz
metastability field and the region where amorphous silica is more
stable that quartz.  The points represented by circles and squares
represent the (vertical) error bars on the pressures needed to
amorphize the system at constant temperature. In the same manner, the
diamonds and triangles are the (horizontal) error bars on the
temperatures needed to melt the system at constant pressure. It can be
noticed that pressurization and heating points give the same results in
the crossover region. The experimental points (crosses) on the
experimental equilibrium quartz, coesit and stishovite melting curves
(thick black line) are reported and the quartz melting curve is then
extrapolated (thick dashed line, star symbols) to higher pressures and
lower temperatures using the measured thermodynamic data for quartz
(data by Zhang {\em et al.} \protect\cite{zhang1}).
The arrow indicates that the extrapolated melting curve must be shifted
to higher pressures in order to account for the configurational entropy
and reduces volume changes due to the presence of five and six-fold
coordinated silicon species \protect\cite{zhang1}.}
\label{phdiag}
\end{figure}

The corresponding points are reported in the P-T phase diagram of
figure~\ref{phdiag}. They represent the numerical coexistence
curve between the crystal, and the {\em melt}.
The term {\em melt} is used here for both the high temperature liquid
state and the high pressure disordered state.  Comparing these points
with the experimental results for the melting of quartz, it seems clear
that the amorphization curve is a reasonable extrapolation of the
melting curve, given that the amorphization process is a continuous
extrapolation of the melting process and falls along the extrapolated melting line of coesite, the highest pressure form of silica consisting of four-fold coordinated silicon clusters. This pressure-induced amorphization line is impossible to measure experimentally, because the amorphization process is impaired by the formation of coesite and stishovite. It represents the metastable (spinodal) stability limit of the quartz phase.
It can furthermore be noted that
the equilibrium melting curve of quartz in the metastable domain
({\em i.e.\/} the coesite stability field) is very close to the experimental
coesite melting curve. This is due to the fact that quartz and coesite
are very similar as far as their local structure is concerned (SiO$_4$
polyhedra), and even though the solid phases have quite different free
enthalpies at low temperature, the latter is smoothed at high
temperature as the system approach the solid--liquid transition curve.
Proof of that is the abscence of any break in the experimental melting
curve of quartz and coesite at the quartz--coesite--liquid triple point.

Calculation of the metastable extension of the equilibrium melting
curve by extrtapolation of thermodynamic data \cite{hemley2} have
improved with the measurement of more accurate data at high temperature
and at high pressure for the SiO$_2$ system \cite{zhang1}.  The recent
calculation by Zhang {\em et al.} \cite{zhang1} using the most recent
thermodynamic data is quite close to the results obtained from the
numerical simulation. Although the extrapolated pressures are lower
that those in the simulation, these authors have argued that these
curves should be shifted to higher pressures if corrected for the
configurational entropy term arising from the distribution of four-
five- and six-coordinated silicon ions in the {\em melt} at high
pressure, and for the volume difference brought by the appearance of
these higher density clusters. In fact, the extrapolation of
thermodynamic data does not include these higher coordination species,
because such a transformation does not occur in the transition from
quartz to coesite.

The continuity between PIA and melting should be reflected into some
similarities between the structural and dynamical properties of the
pressure amorphized material (thereafter called pressure glass) and of
quenched liquids (thermal glasses).  The properties of the latter type
of systems are known to be sensitive to the quenching rate
\cite{vollmayr1}.
In order to compare pressure glasses and thermal glasses, we use as a
reference state a sample of quartz, amorphized at 22~GPa, and then
decompressed to room pressure, leading to a pressure glass with a
density of $\rho=3.07 \; \text{g} \cdot \text{cm}^{-3}$.  Thermal
glasses were produced by cooling a high temperature liquid (7000~K) at
room pressure and constant density $\rho=2.2 \; \text{g} \cdot
\text{cm}^{-3}$ using three different cooling rates $\gamma=4.4 \cdot
10^{12}\; \text{K} \cdot \text{s}^{-1}$, $\gamma=3.5 \cdot 10^{13}\;
\text{K} \cdot \text{s}^{-1}$ and $\gamma=10^{15}\; \text{K} \cdot
\text{s}^{-1}$.
These glassy systems were then compressed at 300~K to obtain densified
glasses with densities of $\rho=3.07 \; \text{g} \cdot \text{cm}
^{-3}$.  A  second set of thermal glasses was  produced  by cooling the
high temperature liquid (7000~K) at a constant density of $\rho=3.07 \;
\text{g} \cdot \text{cm} ^{-3}$, with the same three cooling rates, to
avoid the final compression step in a highly non-ergodic system. Only
the results relative to this last set of samples will be discussed
below; similar, but less precise results were
obtained with the samples cooled at $\rho=2.2 \; \text{g} \cdot
\text{cm}^{-3}$.

\begin{figure}[h]
\begin{center}
\begin{minipage}{1cm}
\hspace*{2mm}\epsfig{file=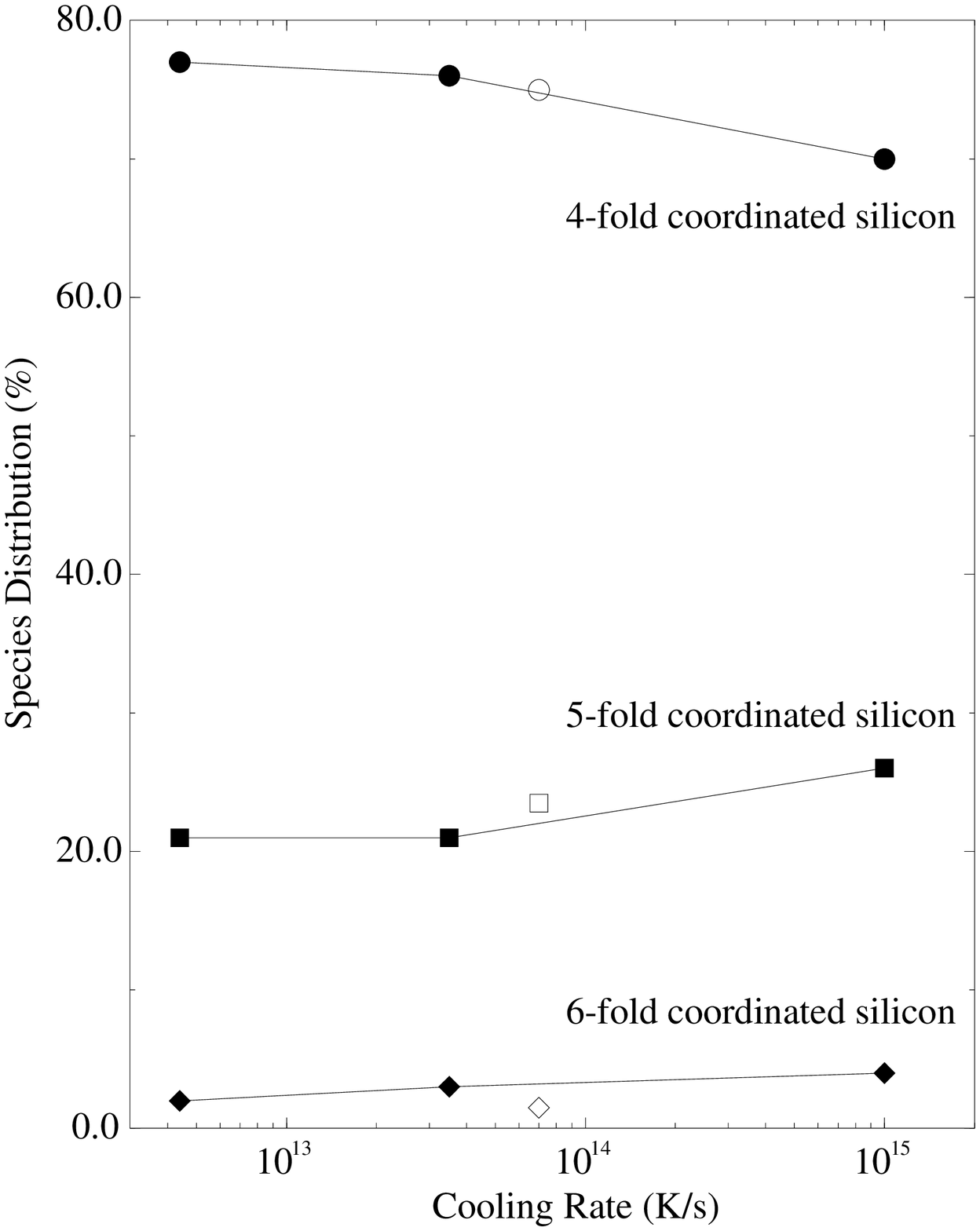,width=6cm}
\end{minipage}
\hfill
\begin{minipage}{1cm}
\hspace*{-5cm} \epsfig{file=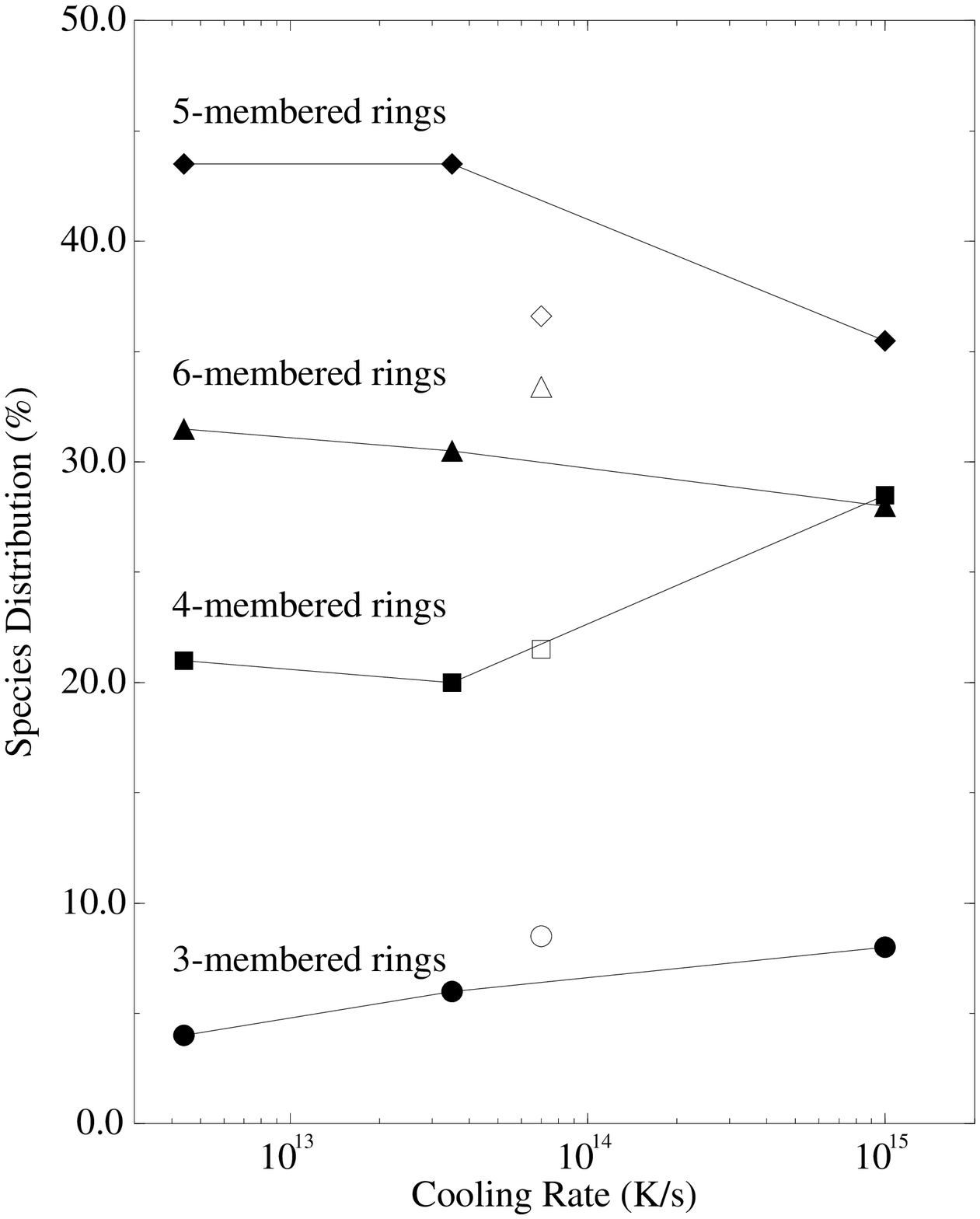,width=6cm}
\end{minipage}
\end{center}
\vspace{15mm}
\caption{The structural properties of the glassy systems at
$\rho=3.07\; \text{g} \cdot \text{cm} ^{-3}$ and 300~K.  The
coordination (left) and ring distributions (right) are reported for the
three thermal glasses (filled symbols) as a function of their cooling
rates.  The corresponding quantities for the pressure glass (open
symbols) are close to what would be obtained for a cooling rate $\gamma
\simeq 7 \cdot 10^{13}\; \text{K} \cdot \text{s}^{-1}$.}
\label{struct}
\end{figure}

\begin{figure}[h]
\centerline{\epsfig{file=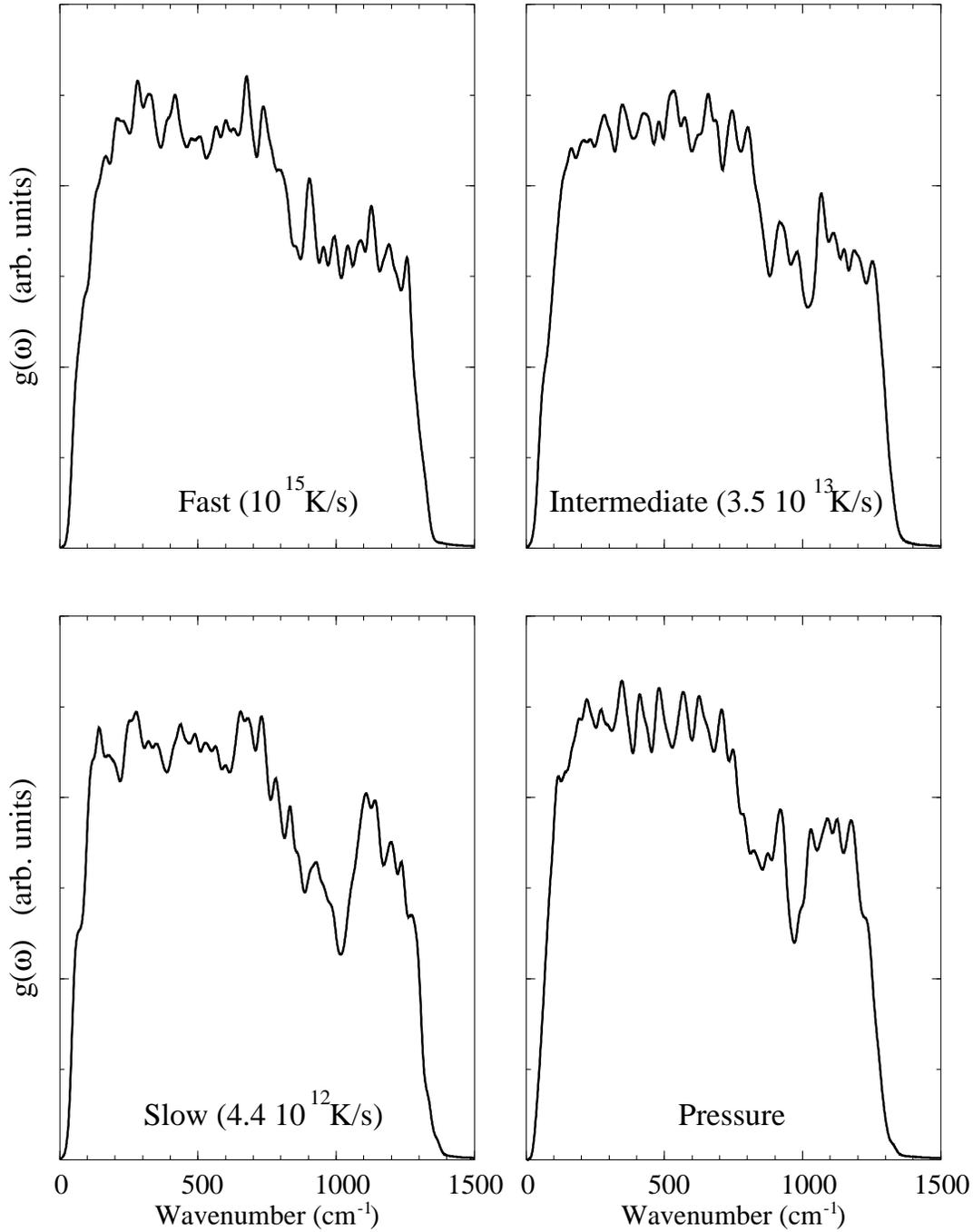,width=14cm}}
\caption{Vibrational densities of states for the glassy systems at
$\rho=3.07\; \text{g} \cdot \text{cm} ^{-3}$ and 300~K.  The three
thermal glasses produced using different cooling rates are reported.
Noticeable differences appear due to non-equilibrium averaging; this
pseudo-dynamical quantity is very sensitive to large scale fluctuations
responsible for frequency shifts in the low frequency region. It can be
noticed that the rapidly cooled systems and the pressure glass are
closer apart, because high-frequency stretching modes of the
tetrahedron are strongly broadened in these cases.  The observation of
another vibrational feature at 900 cm$^{-1}$ leads to the same
conclusion.}
\label{dos}
\end{figure}

Static and semi-dynamical quantities such as the coordination number
distribution, ring statistics (figure~\ref{struct}) and vibrational
densities of state (V-DOS, figure~\ref{dos}) were then calculated for
these systems and compared to the corresponding quantities for the
reference SiO$_2$ pressure glass.

The coordination distribution shows that the proportion of
4-coordinated Si species decreases with increasing cooling rates,
whereas the proportion of 6-fold and especially 5-fold Si coordination
rises.

The effect of the cooling rate on coordination distribution is
consistent with the pressure dependence of the coordination
distribution \cite{barrat2} in liquid SiO$_2$; as pressure rises in the
0--15~GPa range, the proportion of 4-fold coordinated Si species
decreases whereas that of 5-fold (and in a lesser extent to 6-fold)
coordinated Si species increases.  The higher the cooling rate, the
closer are the configurations of the supercooled and high temperature
liquid (higher "fictive temperature").  The high temperature liquid at
these densities is in a high pressure state, which is maintained
during the quench resulting in the formation of a strongly frustrated
amorphous state, because the application of such a high cooling rate
"freezes" the system in such a short time compared with the mechanical
relaxation time that the transformation is quasi-isochoric.
These two observations represent two different pictures of the same phenomenon.

On the other hand, the ring statistics show that a high cooling rate
reduces the proportion of the 5 and 6-membered rings characteristic of
a low density structure in favour of small chains of 3-membered and
4-membered rings.

It has been proposed \cite{hemley3, mcmillan3} that the proportion of
3-membered and 4-membered rings increases with pressure in densified
SiO$_2$ glass or pressure-induced amorphous quartz, on the basis of the
concomitant intensity change of the Raman D$_1$ and D$_2$ defect lines
associated with these rings \cite{galeener1, galeener2}. Our results for the
cooling rate dependence of the ring statistics shows that the
proportion of these clusters is also enhanced by higher cooling rates,
showing that the maximum pressure (before decompression) attained
during room temperature compression of quartz plays a role similar to
that of the cooling rate during the quenching of a melt at room
pressure.

The coordination distribution in the pressure glass is close to that of
a thermal glass obtained with a cooling rate $\gamma \simeq 7 \cdot
10^{13}\; \text{K} \cdot \text{s}^{-1}$ (fig.~\ref{struct}).
The ring distribution displays the same characteristics, except for
6-membered rings whose proportions (within a few percent)
are not very significant, due to the greater error bars in the
determination of these larger scale structures in non-ergodic systems.

A similar conclusion can be drawn from the vibrational density of
states (V-DOS, fig.~\ref{dos}); the high-frequency internal stretching
modes for the SiO$_4$ units are strongly broadened in the pressure
glass. This broadening is an increasing function of the cooling rates in
the thermal glass. Another important
feature is the peak at 900 cm$^{-1}$;
the glasses with higher cooling rates have a narrower peak followed by
a plateau, whereas glasses produced  by slow cooling of the liquid show
a broader peak followed by a drop in the V-DOS.  Once again, the
pressure glass exhibits a behaviour intermediate between that observed
for the two rapidly cooled systems, and is very close to the glass
cooled at $\gamma=3.5 \cdot 10^{13}\; \text{K} \cdot \text{s}^{-1}$.
These results imply that the pressure glass has similar
structural and thermodynamic properties as a rapidly cooled
($10^{13}\; \text{K} \cdot \text{s}^{-1}$ to $10^{14}\; \text{K} \cdot
\text{s}^{-1}$) liquid at the same  density and temperature.  This supports
the continuity between the liquid and the pressure
glass, and therefore between the melting and the amorphization curves.

A natural question is the relationship between our results and the
possibility of a {\em polyamorphic} \cite{poole1, mcmillan2} transition
from a  higher-entropy amorphous form of silica (the pressure glass) to a
lower-entropy form (the thermal glass).  In MD studies, there is no 
indication of any phase transition behaviour during densification,
$\left( \frac{\partial \text{P}}{\partial \text{V}}\right) _{\text{T}}$ seems
negative through the complete densification process at room temperature
\cite{tse1}. In order to accurately check the latter proposition, we
carried out very long (N,P,T) simulations (0.134 nanoseconds) on
SiO$_2$ glass cooled at $\gamma=4.4 \cdot 10^{12}\; \text{K} \cdot
\text{s}^{-1}$ and compressed in the 5--15~GPa range. A histogram of
the density fluctuations was then plotted and it showed a single yet broad
maximum.

\section{Conclusion}

All these arguments suggest that below T$_{\text g}$,
the metastable extension of the quartz melting curve and the
amorphization curve are identical.
non-equilibrium thermodynamic melting can then be
considered as the physical process responsible for high-pressure
amorphization, at least in the case of this theoretical model of silica.

\stars

The calculations were carried out using the CONVEX SPP-1000
computer at the P\^ole Scientifique de Mod\'elisation Num\'erique
(PSMN), \'Ecole Normale Sup\'erieure de Lyon.
We wish to thank Walter Kob for helpful comments and discussions, and
especially for providing us with the initial systems of room pressure
thermal glasses. We thank Bruno Reynard and Russell J. Hemley for helpful comments.

Laboratoire de Sciences de la Terre is UMR 5570. D\'epartement de Physique
des Mat\'eriaux is UMR 5586.

\bibliographystyle{unsrt}

\begin{thebibliography}{10}

\bibitem{mishima2}
O.~Mishima.
\newblock {\em Nature}, {\bf 384}:546--549, 1996.

\bibitem{mishima1}
O.~Mishima, L.D. Calvert, and E.~Whalley.
\newblock {\em Nature}, {\bf 310}:393--395, 1984.

\bibitem{richet2}
P.~Richet and Ph. Gillet.
\newblock {\em Eur. J. Mineral.}, \:in press, 1997.

\bibitem{hemley1}
R.J. Hemley.
\newblock In {\em High-Pressure research in mineral physics}, Mineral Physics
  2. Terra Scientific Publishing Company -- AGU, 1987.

\bibitem{meade1}
C.~Meade, R.J. Hemley, and H.K. Mao.
\newblock {\em Phys. Rev. Lett.}, {\bf 69}:1387--1390, 1992.

\bibitem{itie1}
J.-P. Iti\'e, A.~Polian, G.~Calas, J.~Petiau, A.~Fontaine, and H.~Tolentino.
\newblock {\em Phys. Rev. Lett.}, {\bf 63}:389--401, 1989.

\bibitem{tse1}
J.S. Tse and D.D. Klug.
\newblock {\em Phys. Rev. Lett.}, {\bf 67}:3559, 1991.

\bibitem{tse2}
J.S. Tse and D.D. Klug.
\newblock {\em Science}, {\bf 255}:1559--1561, 1992.

\bibitem{chelikowsky1}
N.~Binggeli, N.~Troullier, J.-L. Martins, and J.R. Chelikowsky.
\newblock {\em Phys. Rev. B}, {\bf 44}:4471, 1991.

\bibitem{binggeli3}
N.~Binggeli and J.R. Chelikowsky.
\newblock {\em Phys. Rev. Lett.}, {\bf 69}:2220--2223, 1992.

\bibitem{mcneil1}
L.E. McNiel and M.~Grimsditch.
\newblock {\em Phys. Rev. Lett.}, {\bf 68}:83--85, 1992.

\bibitem{watson1}
G.W. Watson and S.C. Parker.
\newblock {\em Philosophical Mag. Lett.}, {\bf 71}:59--64, 1995.

\bibitem{binggeli97}
N. Binggeli.
\newblock {\em Simulation of Silicas : from classical pair potentials
to density functional theory}, {CECAM Workshop} : September 15--17,
1997.

\bibitem{badro96}
J.~Badro, J.-L. Barrat, and Ph. Gillet.
\newblock {\em Phys. Rev. Lett.}, {\bf 76}:772--775, 1996.

\bibitem{hemley2}
R.J. Hemley, A.P. Jephcoat, H.K. Mao, L.C. Ming, and M.H. Manghnani.
\newblock {\em Nature}, {\bf 334}:52--54, 1988.

\bibitem{richet1}
P.~Richet.
\newblock {\em Nature}, {\bf 331}:56--58, 1988.

\bibitem{beest1}
B.W.H. {van Beest}, G.J. Kramer, and R.A. {van Santen}.
\newblock {\em Phys. Rev. Lett.}, {\bf 64}:1955--1958, 1990.

\bibitem{zhang1}
J.~Zhang, R.C. Liebermann, T.~Gasparik, and C.T. Herzberg.
\newblock {\em J. Geophys. Research}, {\bf 98}:19785--19793, 1993.

\bibitem{vollmayr1}
K.~Vollmayr, W.~Kob, and K.~Binder.
\newblock {\em Phys. Rev. B}, {\bf 54}:15808--15827, 1996.

\bibitem{barrat2}
J.-L. Barrat, J.~Badro, and Ph. Gillet.
\newblock {\em J. Mol. Sim.}, {\bf 20}:17--25, 1997.

\bibitem{hemley3}
R.J. Hemley, H.K. Mao, P.M. Bell, and B.O. Mysen.
\newblock {\em Phys. Rev. Lett.}, {\bf 57}:747--750, 1986.

\bibitem{mcmillan3}
P.F. McMillan, B.T. Poe, Ph. Gillet, and B.~Reynard.
\newblock {\em Geochim. Cosmochim. Acta}, {\bf 58}:3653--3664, 1994.

\bibitem{galeener1}
R.J. Bell, N.F. Bird and P. Dean.
\newblock {\em J. Phys. C}, {\bf 1}:299, 1968.

\bibitem{galeener2}
F.L. Galeener, A.J. Leadbetter and M.W. Stringfellow.
\newblock {\em Phys. Rev. B}, {\bf 27}:1052, 1983.

\bibitem{poole1}
P.H. Poole, T.~Grande, F.~Sciortino, H.E. Stanley, and C.A. Angell.
\newblock {\em Comp. Mat. Sci.}, {\bf 223}:1--9, 1995.

\bibitem{mcmillan2}
S.~Aasland and P.F. McMillan.
\newblock {\em Nature}, {\bf 369}:663, 1994.

\bibitem{gillet90}
Ph.~Gillet, A.~Le~Cl\'eac'h, and M.~Madon.
\newblock {\em J. Geophys. Res.}, {\bf 95}:21635--21655, 1990.

\bibitem{grimsditch94}
M. Grimsditch, S. Popova, V.V. Brazhkin, and R.N. Voloshin.
\newblock {\em Phys. Rev. B}, {\bf 50}:12984--12986, 1994.

\end{thebibliography}

\newpage

\vspace*{1cm}

\newpage

\vspace*{3cm}

\newpage

\end{document}